\documentstyle[referee]{mn}

\title{Tidal and Tidal-Resonant Effects in Coalescing Binaries}

\author[ Kostas D. Kokkotas and Gerhard Sch\"afer ]
{Kostas D. Kokkotas$^{1,2}$ and Gerhard Sch\"afer$^{1}$ \\
$^1$Max-Planck-Gesellschaft, Arbeitsgruppe Gravitationstheorie an der
Friedrich-Schiller-Universit\"at Jena, 07743 Jena, Germany \\
$^2$Department of Physics, Aristotle University of Thessaloniki,
    540 06 Thessaloniki, Macedonia, Greece.
}

\begin{document}

\maketitle
\date{Received }
\begin{abstract}
Tidal and tidal-resonant effects in coalescing compact binary systems are
investigated by direct numerical integration of the equations of motion.
For the stars polytropic models are used. The tidal effects are found
to be dominated by the (non-resonant) $f$-modes. The effect of the
$g$-mode-tidal resonances is obtained. The tidal interaction is shown to be
of interest especially for low-mass binaries.
There exists a characteristic
final plunge orbit beyond which the system cannot remain
stable even if radiation reaction is not taken into account; in
agreement with results obtained by Lai et al. \shortcite{Lai93}.
The importance of the investigated effects for the
observation of gravitational waves on Earth is discussed.
\end{abstract}

\section{ Introduction}

Close neutron star binaries are among the primary sources of
gravitational waves that will be detected by the future
gravitational-wave observatories on Earth \cite{Th87}.
Using the matched filtering technique proposed by
Thorne \shortcite{Th87}, close binaries should be detectable out to
cosmological distances in the Gpc regime.  The observed
gravitational waves will be produced
during the final inspiral and coalescence stages of the motion of the both
compact objects (neutron stars and/or black holes). The detection process
will last 2 -- 3 secs as the gravitational waves are expected to
be measured when the binaries will be in the gravitational-wave-frequency
window of 100 -- 1000 Hz.
Future improved detector technology is expected to widen this window
down to 10 Hz and up to 10 kHz, which means longer observation time
and higher resolution of the signal, leading in this way to a more accurate
determination of the system parameters like the masses, spins, and radii of
the bodies. Nevertheless, already the 100 -- 1000 Hz window can give us
important information about the above parameters of the system if
we know the orbital motion and the interaction mechanism
of the two extended bodies with reasonable accuracy (during the final
inspiral the bodies cannot be considered as point-like).
The better our theoretical knowledge will be the more information will be
extractable from the detected signal while any effect that will be omitted
will reduce the signal to noise ratio of the detector.
For example, the omission of post-Newtonian orbital terms in the
construction of the template waveform will not change significantly
its amplitude but it will produce destructive effects on the
phase of the wave leading to a decrease in the correlation
of the incoming wave and the applied template \cite{BFS93}.
Recent numerical simulations
for testing the post-Newtonian approximation \cite{KKT94} have shown that the
correlation falls down by $\approx 40\%$ for nearly equal-mass systems while
the phase effects become more dramatic for systems with unequal
masses. Moreover, this phase mismatch will produce
analogous effects in the accuracy of the calculation of the system
parameters (e.g. masses) as well
as of the observational distance and, as a consequence, of the rate of the
detected events \cite{KKT94}. For these reasons a lot of attention
has been paid recently to the study of possible
fluid characteristics that may corrupt the signal \cite{Koc92,JK92,BC92,SK94}.

What is expected from theoreticians in the remaining 5 to 6 years before
operating gravitational--wave observatories might be available
is the qualitative and quantitative study of every aspect of the compact
binary--star coalescence in order to construct the best possible theoretical
models that our theoretical approximations allow \cite{Calt92,BFS93}.
Tidal effects have been proposed \cite{FPR75} as the reason for the
formation of close binary systems. Tidal capture has been extensively
studied by various authors in the past two decades leading to a quite
well understanding of the influence of the tides not only on the formation
but also on the evolution of close binary systems \cite{PT77,LO86,Ale87}.

In this line of research we are investigating tidal and tidal-resonant
effects in the final stages of the inspiral of two neutron stars. Beyond this
we investigate the information that can be gained about these effects
from the detected gravitational waveforms. We also calculate the evolution
of the binaries in the 1st-post-Newtonian (1PN) approximation to general
relativity, including 3.5PN gravitational radiation damping, and compare
it with the tidal evolution.

In the next section we present equations that govern the evolution
of close binary systems under the influence of Newtonian tidal and
2.5PN gravitational (quadrupole-)radiation reaction forces. In the third
section we describe numerical procedures and discuss the obtained results.

\section {Equations of Motion }

Let us consider a close binary system consisting of two bulk-non-rotating
neutron stars with
masses $m_1$ and $m_2$. Both of the stars are more or less tidally
distorted by their respective companions.
The Hamiltonian governing the motion of the two extended bodies, in our case,
consists of four parts: the contribution from the stationary orbital motion,
a part describing the oscillations of the stars, a term
arising from the gravitational interaction of the tidally deformed star 2
with the undeformed star 1, and vice-versa, and, finally, a part
which describes the gravitational radiation
reaction or radiation damping.

The orbital Hamiltonian, at Newtonian order, in polar coordinates reads
\begin{eqnarray}
H_{orb}  =  {1\over{2\mu}} \left(P_R^2 + {{P_\Phi^2} \over R^2}  \right)
         -  { {G \mu M}\over R},
\end{eqnarray}
where $R$ denotes the separation of the two stars and $\Phi$ the orbital
rotation angle. $P_R$ and $P_\Phi$ are the
corresponding canonical momenta.
\begin{equation}
\mu = {{m_1 m_2}\over{m_1+m_2}} \quad \mbox{and} \quad  M = m_1 + m_2
\end{equation}
are the reduced mass and the total mass of the system, respectively.

The Hamiltonian describing the oscillations of the stars corresponds to
an infinite set of harmonic oscillators. It has the
form \cite{Ale87}:
\begin{equation}
H_{osc}^{(a)} = {1\over 2} \sum_n \left( p_n^2 + \omega_n^2 q_n^2 \right),
\end{equation}
where ${q_n}$ are generalized coordinates and ${p_n}$ their conjugate
momenta; $\omega_n$ are real frequency eigenvalues. The index $a$ on
$H_{osc}$ denotes the stars; it takes the values 1 and 2. Like in eq. (3)
we often drop indices for simplicity.

The tidal gravitational potential at a point $(r,\theta,\phi)$ inside, let us
say, the star 2 (the centre of mass of this star is put into the origin of the
coordinate system) due to the star 1, which is located at
$(R,{\pi}/2,\Phi)$ and treated as point-like, reads in spherical coordinates:
\begin{equation}
\phi_T = - {{4 \pi G m_1}\over R}
\sum_{\ell=2}^{\infty} \sum_{m=-\ell}^\ell {1\over { 2 \ell + 1}}
Y_{\ell m}({\pi/2},\Phi) Y^*_{\ell m}(\theta,\phi) (r/R)^{\ell},
\end{equation}
where $Y_{\ell m}$ are normalized spherical harmonics.
Using this expression, the gravitational potential energy $U_T$ of
the deformed star 2 (with the corresponding mass change $\delta \rho$)
in the tidal field of star 1 is given by
\begin{equation}
U_T^{(1,2)} = \int \phi_T \delta \rho d^3 r .
\end{equation}

By the use of the continuity equation ($\rho_0$ is the unperturbed mass
density)
\begin{equation}
\delta \rho = - {\rm div} \rho_0 \mbox{\boldmath$\xi$}
\end{equation}
and integration by parts, the expression (5) becomes
\begin{equation}
U_T^{(1,2)} = \int \rho_0 \mbox{\boldmath$\xi$} \cdot \nabla \phi_T d^3 r ,
\end{equation}
where \boldmath$\xi~$\unboldmath is a vector describing the Lagrangian
displacement of the fluid. The vector can be written as a linear superposition
of normalized eigenvectors  $\mbox{\boldmath$\xi$}_n$,
\begin{equation}
\mbox{\boldmath$\xi$}({\bf r},t) = \sum_n q_n(t) \mbox{\boldmath$\xi$}_n
({\bf r}),
\end{equation}
which satisfy a self-adjoint eigenvalue equation of the type
\begin{equation}
{\cal L} \mbox{\boldmath$\xi$}_n = \rho_0 \omega_n^2 \mbox{\boldmath$\xi$}_n.
\end{equation}

The normal modes $\mbox{\boldmath$\xi$}_n$
are characterized by their spherical harmonic
indices $\ell$ and $m$, and the radial index $n$.
They can be decomposed
into radial and poloidal spherical harmonics \cite{Cha64}:
\begin{equation}
\mbox{\boldmath$\xi$}_n \equiv \mbox{\boldmath$\xi$}_{n \ell m}
(r,\theta,\phi) =
\left[ {\xi}^R_{n \ell}(r) {\bf e}_r +
{\xi}^S_{n \ell} (r)r \nabla \right] Y_{\ell m}(\theta,\phi).
\end{equation}
By the aid of the unitary transformation
\begin{eqnarray}
\mbox{\boldmath$\xi$}_{n\ell m}^{(e)}
   & = & {1\over \sqrt{2}}\left[{\mbox{\boldmath$\xi$}}_{n\ell m}
  + {\mbox{\boldmath$\xi$}}_{n\ell -m} \right] \qquad (0 \leq m \leq \ell), \\
\mbox{\boldmath$\xi$}_{n\ell m}^{(o)}
   & = & {1\over {{\rm i} \sqrt{2}}}\left[{\mbox{\boldmath$\xi$}}_{n\ell m}
     - {\mbox{\boldmath$\xi$}}_{n\ell -m} \right] \qquad (1 \leq m \leq \ell)
\end{eqnarray}
we can create an orthonormal real basis
in which the generalized coordinates ${q^{(\sigma)}_n}$ ($ \sigma = e, o$)
are real. Herewith
$U_T$ can be written as
\begin{equation}
U_T^{(1,2)} = -  {{G m_1 m_2}\over R}
\sum_n \sum_{\ell = 2}^{\infty}
\sum_{\sigma = e,o} \sum_{m = 0}^{\ell}\prime W_{\ell m}
{{q_{n \ell m}^{(\sigma)}} \over r_2} \left( {{r_2} \over R} \right)^{\ell}
Q_{ n \ell} 2^{1/2}\mbox{trig}^{(\sigma)}_{m}(\Phi) ,
\end{equation}
where $\mbox{trig}^{(e)}_{m}(\Phi) = \mbox{cos}(m \Phi)$,
$\mbox{trig}^{(o)}_{m} (\Phi) =
\mbox{sin}(m \Phi)$.
 $\Phi$ decomposes into $\Phi = v + g$, where $v$ is the true
orbital anomaly and $g = \mbox{constant}$.
The prime on the fourth sum means exclusion of the case
$m=0,\sigma=o$, and $r_2$ is the radius of the star 2. $Q_{n\ell}$
denotes the overlap integral
\begin{equation}
Q_{n\ell} = \ell \int_0^1 {\bar \rho} {\bar r}^{\ell+1}
\left[ \xi^R_{n\ell}({\bar r})+(\ell+1)\xi^S_{n\ell}({\bar r}) \right]
d {\bar r},
\end{equation}
where $\bar r = r/r_2$ and $\bar \rho = \rho_0 r_2^3/m_2$
is a dimensionless density and $W_{\ell m}$ is given by
\begin{equation}
W_{\ell m}~=~(-)^{\frac{\ell+m}{2}}\left[\frac{4\pi}{2\ell+1}(\ell-m)\mbox{!}
(\ell+m)\mbox{!}\right]^{\frac{1}{2}} {\bigg/}
\left[2^{\ell}\left(\frac{\ell+m}{2}
\right)\mbox{!}\left(\frac{\ell-m}{2}\right)\mbox{!}\right] ,
\end{equation}
where $(-)^k$ is zero if k is not an integer. The functions
$q^{(\sigma)}_{n \ell m}$ of the motions of the harmonic oscillator, defined
by the Hamiltonian
(3), in terms of action--angle variables, $J^{(\sigma)}_{n \ell m}$ and
$\theta^{(\sigma)}_{n \ell m}$, respectively, can be represented as
\begin{equation}
q^{(\sigma)}_{n \ell m} = (2 J^{(\sigma)}_{n \ell m} / \omega_{n \ell})^{1/2}
\mbox{cos} \theta^{(\sigma)}_{n \ell m} ,
\end{equation}
where $\theta^{(\sigma)}_{n \ell m} = \omega_{n \ell} t +
\phi^{(\sigma)}_{n \ell m}$,  $\phi^{(\sigma)}_{n \ell m} = \mbox{constant}$.
Herewith, finally, $U_T^{(1,2)}$ takes the form
\begin{eqnarray}
U_T^{(1,2)} = &-& {{G m_1 m_2}\over R}
\sum_n \sum_{\ell = 2}^{\infty}
\sum_{\sigma = e,o} \sum_{m = 0}^{\ell}\prime  W_{\ell m}
{1 \over r_2} \left( {{r_2} \over R} \right)^{\ell}
Q_{ n \ell} \omega_{n \ell}^{-1/2} \nonumber\\
&\times& \left[(J^{(e)}_{n \ell m})^{1/2} \{\cos(\theta^{(e)}_{n \ell m}
+ m \Phi)
+ \cos(- \theta^{(e)}_{n \ell m} + m \Phi) \} \right.\nonumber\\
&+& \left.(J^{(o)}_{n \ell m})^{1/2} \{\sin(\theta^{(o)}_{n \ell m}
+ m \Phi) + \sin(- \theta^{(o)}_{n \ell m} + m \Phi) \}\right].
\end{eqnarray}

Circular orbits play an important role in coalescing binaries. There,
$\dot \Phi$ is given by the single Keplerian orbital angular frequency
$\omega_K$. The orbital motion induces perturbations on the
star modes which rotate with the phase velocities $m \omega_K$, see eq.\ (13).
For an adiabatic inspiral situation $\omega_K$ is
a slowly varying function with time. Whenever $\omega_{n \ell} / \omega_K$
approaches integer values, resonances can occur; see eq.\ (17) and,
for $W_{\ell m}$, remember the condition $(\ell + m)/2 = \mbox{integer}$.

The Hamiltonian of the entire system  --  orbital
motion, star oscillations, tidal coupling of the two stars, radiation reaction
  --  takes the form:
\begin{equation}
H(t) = H_{orb} + H_{osc}^{(1)} + H_{osc}^{(2)} + H_{tid}^{(1,2)} +
H_{tid}^{(2,1)} + H_{reac}(t) ,
\end{equation}
where $H_{tid}^{(1,2)} = U_T^{(1,2)}$, etc.

The (explicitly) time-dependent reaction Hamiltonian $H_{reac}$ is given by,
cf.~\cite{Sch90}:
\begin{equation}
H_{reac}(t) = \frac{2G}{5c^5} \frac{d^3Q_{ij}(t)}{dt^3}
\left(\frac{P_iP_j}{\mu} - GM\mu \frac{R^iR^j}{R^3} \right),
\end{equation}
where $Q_{ij} = \mu \left(R^iR^j - \frac{1}{3} \delta_{ij}R^2 \right)$
denotes the mass-quadrupole tensor of the two-body system.

A spin-orbit-interaction part $H_{SO}$ could have been also included
into the Hamiltonian, e.g. see \cite{DS88}. But it has been
found that in our case of bodies without bulk rotation it does not contribute
significantly to the evolution of the system, and in what follows we shall
not discuss it any further.

The system evolves in time as dictated by the Hamiltonian equations of
motion:
\begin{eqnarray}
{d {\bf P} \over {d t}} & =& - {{\partial H(t)}\over {\partial {\bf R}}},
\quad
{d {\bf R} \over {d t}}   =  + {{\partial H(t)}\over {\partial {\bf P}}}, \\
{d {p_n} \over {d t}} & =& - {{\partial H(t)}\over{\partial {q_n}}} , \quad
{d {q_n} \over {d t}}   = + {{\partial H(t)}\over{\partial {p_n}}}.
\end{eqnarray}

The respective reaction parts, ${\bf F}_{\bf P} $ and ${\bf F}_{\bf R}$,
in the above equations, in polar coordinates, take the form, cf.~\cite{Sch85}:
\begin{eqnarray}
F_{P_R} &=& +{8\over 3} {{G^2 P_R}\over{c^5 R^4}}
\left({{G M^3 \nu}\over 5} - {{P_\Phi^2} \over {\nu R}} \right) , \\
F_{P_\Phi} & =& -{8\over 5} {{G^2 P_\Phi}\over{c^5 R^3 \nu}}
\left({{2 G M^3 \nu^2}\over R} + 2 {{P_\Phi^2} \over {R^2}}
- P_R^2 \right) , \\
F_{R} &=& -{8\over {15}} {{G^2 }\over{c^5 R^2 \nu}}
\left( 2 P_R^2  + 6 {{P_\Phi^2} \over {R^2}} \right) , \\
F_{\Phi} &=& -{8\over 3} {{G^2 P_R P_\Phi}\over{c^5 \nu R^4}}.
\end{eqnarray}

The gravitational waves are emitted primarily by the orbital motion.
The amplitudes of the leading (quadrupole) waves are proportional
to $1/R$ and their phases vary with time through $2 \Phi$.


\section{Results and Discussion}

The Hamiltonian equations of motion (20) -- (21) have been integrated
for various
binary systems consisting of neutron stars from the less relativistic
to the extreme relativistic regime.
The equation of state was taken to be of polytropic type,
\begin{equation}
p = K \rho^\Gamma , \quad \mbox{where} \quad \Gamma=1+{1\over N}
\end{equation}
with polytropic index $N=0.5, 1, 2$ while the central density was
$\approx 10^{14} - 10^{15} \mbox{gr/cm}^3$.
It is known that the best approximation
for constructing realistic polytropic neutron stars is to take polytropes
with indices 0.5 up to 2 \cite{Finn87}.
The adiabatic index $\Gamma_1$ we have chosen in such a way that $g$-modes
exist. This is possible as far as the adiabatic index $\Gamma_1$ is larger than
$\Gamma$; in the limit $\Gamma_1 = \Gamma$ there are no g-modes. There are two
other important aspects regarding the selection of the appropriate adiabatic
index: As $\Gamma_1 \to \Gamma$ the g-mode frequencies
are slowing down significantly and the g-modes become degenerate at
zero frequency. This property of the g-modes in principle increases the
influence of the interaction part of the Hamiltonian (since $U_T \approx
\omega^{-1/2}$) but there are also the values of the overlap integrals which
decrease much faster as it can be seen from Tables 2a and 2b.
The result is that
the interaction Hamiltonian becomes less important as $\Gamma_1 \to \Gamma$.
For a detailed discussion of the g-modes we refer to Finn \shortcite{Finn87},
McDermott et al. \shortcite{McDer85},
and Reisenegger and Goldreich \shortcite{ReGol92} and references therein.
For the $f$-modes it does practically not matter
which values of $\Gamma_1$ we use, the frequency and the
overlap integral are not very sensitive to the choice of $\Gamma_1$.
In what follows we shall discuss polytropic stars which show typical external
characteristics, i.e. masses and radii of accepted relativistic neutron stars
models. More general equations of state will surely alter
the results, but the main outcome of the paper,
in particular those effects which are dominated by the $f$-modes,
should have wider applicability.
We have decided to study the binary evolution for nine
different star models, A, B, C with polytropic indices $0.5$, $1$, and $2$
(the value of the constant K has been chosen appropriately in each case).
These models span the whole range of the
neutron star family, from the very condensed state with high surface
redshift to the less compact state with small surface redshift, see Table 1.
In all cases the equations of structure are Newtonian-like, i.e. $\rho$
in eq.\ (26) is the Newtonian mass density.
Thus, with the models we have constructed, the
less relativistic neutron stars are the better approximated ones.
For these models normal-mode eigenfrequencies and eigenfunctions have been
found and overlap integrals have been calculated. For $\ell = 2$, the
part of the spectrum
around the $f$-mode together with the first few {\it gravity} and
{\it pressure} modes for the $N=0.5, 1, 2$ polytropes
are shown in Tables 2a and 2b
where additionally the corresponding values of the overlap
integrals are listed.
Although for our numerical calculations we have mainly used the values of
Table 2a we
present also Table 2b in order to show the dependence of the g-mode
frequencies and the operlaping integrals on the values of the adiabatic index.
The frequencies agree with those given by Cox \shortcite{Cox80}.
For an easier use of Tables 2a and 2b, in the last column of Table 1, the
fundamental frequencies are given in units of Hz. When we
checked the procedure for the calculation of the overlap integrals
we have calculated also the normal modes and the overlap integrals for
$N=3$ polytropes ($\Gamma_1 = 5/3$). The agreement with the results of
Lee \& Ostriker \shortcite{LO86} was very good, usually the difference was less
than $0.1\%$ even for the lowest $g$-modes for which special care
must be taken in order to avoid numerical errors.

The procedure followed for the calculation of the normal-mode
eigenfrequencies
is similar to Dziembowski \shortcite{Dzi71} with the exception that the final
frequencies and
eigenfunctions calculated by the aid of this procedure have been used as trial
expressions for a variational calculation of the eigenfrequencies.
Practically, this final step together with an extra check
for the orthogonality
of the eigenfunctions was just a verification for the accuracy
of our method.

The Hamiltonian equations of motion (20) -- (21), for each value of
$n$, $\ell$, $m$ ($m$ positive), and $\sigma$, form a system of 6
first-order ODEs. As can be seen from
the form of $U_T$ (see eq. (17)), because of the existence of the term
$(r_2/R)^{\ell}$, respectively $(r_1/R)^{\ell}$,
only the harmonics with $\ell=2$ and $3$ are
important ($m=0,\dots,\ell$). Furthermore, from the Tables 2a and 2b
follows that the $f$-mode together with the first
$g$- and $p$-modes maximize the overlap integral, but since $U_T$ is
inverse proportional to the square root of $\omega_{n\ell}$ the $p$-modes
do not contribute significantly and thus their incorporation in the following
calculations is not needed.

The initial conditions put two non-oscillating stars into circular orbits
with orbital frequencies of $50Hz$ (correspondingly, the frequencies
of the emitted gravitational waves are $100Hz$). Notice that at the late stages
of the evolution of a
binary system orbits are expected to be circular because of the tendency of
the gravitational radiation damping to circularize the orbits.
Then the system is left to evolve
according to the equations (20) -- (21). The evolution of the system depends
on the radiation damping and on the tidal interaction of the two stars.

Since the important aspect of this work is the comparison of the rate of
evolution of binary systems with and without tidal interaction we have to
discuss also a refinement of the binary motion where the components are
treated as point-like. For this we consider the 1PN orbital dynamics.
In case of circular motion the (secular) radiation-reaction change of
the 1PN orbital angular frequency reads, e.g see \cite{CF94}
\begin{equation}
{{d \omega}\over {d t}} = {{96 G^{5/3} \mu M^{2/3}}\over {5 c^5}}
\mu M^{2/3} \omega^{11/3} \left[ 1 - {{(743+924 \mu / M)}\over
{336 c^2}}(G M \omega)^{2/3}  \right].
\end{equation}
In this relation the gravitational radiation damping has been taken into
account up to the 3.5PN level (for non-circular orbits, see \cite{JS92})
using balance equations.
A local expression for the 3.5PN reaction force is still not available.

As seen from Figure 1, the evolution of the system as calculated
by the damped 1PN approximation (equation 27) is slower
than the evolution which assumes damped Newtonian approximation.
If we additionally turn on the tidal part of the Hamiltonian, $H_{tid}$, then
the evolution becomes faster since the orbital motion will excite
oscillations of the initially non-oscillating stars. The amount of the energy
transferred from
the orbital dynamics to the stellar dynamics depends on the stellar
equation of state and the excitation efficiency of the tidal coupling.
As mentioned earlier, the overlap integral $Q_{n2}$ is maximum
for the $f$-mode, and it has been shown numerically that
all other modes together cannot absorb more
orbital energy than the $f$-mode alone (see also \cite{reisen94}).
If radiation reaction is turned off, the tidal interaction makes the orbits
oscillating around the initially chosen circular orbit. Without radiation
damping the system is a near-integrable system since the part of the
Hamiltonian which describes the orbital motion and the oscillations of the star
is integrable while the other part which describes the tidal interaction
is the ``perturbation Hamiltonian''. Therefore, techniques for the treatment
of evolution and resonances of near-integrable systems can be applied
\cite{Ale87,Gol80,LL83}. For an extensive analysis of the
resonances in Newtonian binary systems the reader is referred to a paper by
Alexander \shortcite{Ale87}.

For the cases we are studying, i.e. coalescing binary systems
consisting of two
neutron stars, strong $f$-mode resonances
can occur only when the two stars are near their final plunge stage.
Since the orbit there shrinks very fast, the passage through
the tidal resonances is instantaneous and thus it does not result in a
significant energy exchange between the quasi-periodic orbital part and the
periodic internal part of the system.
But if we allow for the lower frequency $g$-modes then resonances can occur
as it is shown in Figures 2 and 3. Obviously, for $\omega_{n2} = 2 \omega_K$,
i.e. $\ell = 2$ and $m = 2$, the tidal-interaction Hamiltonian leads to
strongest
resonance, see eq.\ (17), and the $g$-mode frequencies are low enough
to fulfil this condition, see Table 2. However, the Figures 2 and 3 show
that the $f$-mode (non-resonant) effects are much more important than
$g$-modes resonant ones.
The effect of the resonances depend on the adiabatic index. We find
that the g-mode resonances are unimportant for the polytropes with
$N=0.5$ but have some relevant contribution as $\Gamma_1$ becomes
larger which is mainly due to the small values of the overlapping integrals.
The inverse is true for the f-modes. The f-mode frequency
for the case $N=0.5$ is smaller and it results that the deviation
from the point-particle evolution is larger, see Figures 1 to 4.

In Figures 2 and 3 one can see the differences between
the distances of the two stars
as calculated with radiation reaction + tidal interaction on the one side
and with radiation reaction alone on the other.
The differences start oscillating as the orbital motion of the
two stars approaches frequencies of one half of the oscillation frequencies.
The figures show the effects of the $f$-, $g_1$-, and $g_2$-modes for the
binary systems $A_{0.5}-C_{0.5}$ and $A_1-C_1$.
Both figures clearly show that the $f$-modes dominate the
$g$-modes. The effects of the tidal effects
mildly corrupt the nice sinusoidal form that the standard matched filtering
technique is supposing. This leads to a smaller correlation and results in
an increase of the errors in the calculation of the mass parameters
of binaries from measured waveforms. Evidently, near the high--frequency
end of the gravitational-wave-observation window tidal effects
should be incorporated into the template waveform. This is supported also by
the recent works by Reisenegger and Goldreich \shortcite{ReGol94} and
Shibata \shortcite{Shib94}.

For the total phase, see Figure 4,
we have obtained a phase difference between the (orbital)
motion with radiation reaction and the motion with radiation reaction +
tidal interaction of the order $15 \pi$ for a binary consisting of two
stars of type A$_1$-C$_1$ (see Table 1) and $21 \pi$ for a binary consisting
of one type A$_{0.5}$ and one type C$_{0.5}$ star,
while in the case of a binary with
stars of types A$_i$ and A$_i$ the phase difference is quite small and
unimportant. In both cases these phases have been
accumulated during the 100 -- 1000 Hz emission stage of gravitational radiation
while the stars have completed about 300 revolutions around each other.

While the tidal resonances contribute seemingly only during the late
stages of the binary-systems evolution,
the tidal effects accumulate energy on the stellar oscillations
during the whole period of the coalescence.  The transfer of energy
becomes significant if the system has the following characteristics:
(i) long coalescing time, which implies small stellar masses
since the coalescing time is proportional to $\approx M^{-3}$ for
nearly equal-mass binaries, and
(ii) small ratios $m_{i}/r_{i}$, because in this case the tidal effects
are more significant (it results in a larger energy transfer from
the orbital motion to
the oscillations). Even though, the contribution of the tidal effects
outside the region where resonances get important
is in general small. For the system A$_1$-C$_1$ we have mentioned
earlier that during the whole evolution of the system, from the
beginning of the observation window, i.e. 100Hz which correspond
to a distance of $44.8 M$, the phase difference accumulates to $14.2 \pi$,
while the evolution of the same system from a distance of $20 M$
(frequency of the gravitational waves 335 Hz) to $9.7 M$ (end of the
observational window, where also the stars touch each other)
accumulates a phase difference of $12.9 \pi$. The corresponding numbers for
the system A$_{0.5}$-C$_{0.5}$ are $21 \pi$ and $18.3\pi$.
This means that $90\%$
of the total phase difference is created during the final inspiral.
This late-stages corruption of the signal due to the tidal
effects is also apparent in the work of Kochanek \shortcite{Koc92}
where irrotational Roche-Riemann ellipsoids have been used as models
for the neutron stars.

An interesting feature is the fact that if the two stars are close enough the
tidal effects can drive them into coalescence even in the case that
no radiation reaction is taken into account, see Figure 5.
The {\sl tidally induced plunge} region strongly depends on the masses of
the two stars and on their degree of compactness. The
more compact the stars are the smaller their separation is for the
tidally induced plunge. For the three models that we referred to earlier
this distance takes the values:
$11.6 M$ ($8.9 M$) for the system A$_2$-C$_2$,  $16 M$ ($8.9 M$) for the
system A$_1$-C$_1$, $17.4 M$ ($8.9 M$) for the system A$_{0.5}$-C$_{0.5}$
and  $19 M$ ($15.9 M$) for the system C$_1$-C$_1$ (in
parentheses are shown the distances for which the stars touch each other).
We should notice in addition that this tidally induced plunge, in reality,
should be more
complicated since it happens for distances where the two stars have passed
already their Roche limit \cite{BC92}. Recently Lai et al. \shortcite{Lai93},
\shortcite{Lai94} came to similar conclusions for binary systems with also
polytropic star components, but they have not used the approach which is
applied in the present work. The results support the
idea of introducing the notion of extended bodies in the construction of
the waveform templates for the future detection of gravitational waves.
Nevertheless, we should keep in mind that we have used linear
analysis and that the ``plunge orbit" occurs at
small separations at which the linear theory is not strictly valid. Thus,
although the existence of ``plunge orbits" seems to be established the
numbers that we give might change if non-linear theory is
applied.

Concluding this work we will discuss the effects of our models
on the waveform of the gravitational waves emitted during the
binary coalescence stage.
As it can be seen from the Figure 4 (remember the end of Sect.\ 2) the waves
accumulate very quickly a phase difference which seriously
corrupts the shape of the Newtonian waveform leading to difficulties
in the detection procedure if not taken into account in the
template waveform (for a numerical simulation of the detection
procedure and the exact influence on the detection procedure the reader is
referred to \cite{KKT94}). Although the 1PN corrections
are weakening the high phase differences \cite{KKS94} the effects on the
phase are very large and cannot be overcome.
For the evolution of the system A$_i$-C$_i$,
from $20 M$ to coalescence, the phase difference produced by the
1PN  theory is $32 \pi$,
while from the tidal effects, as mentioned earlier, $21 \pi$ for the system
A$_{0.5}$-C$_{0.5}$ and $14 \pi$ for the system A$_1$-C$_1$.
The large tidal contributions
depend on the choice of the binary system. If we would have chosen
a system of the type A$_i$-A$_i$ then the post-Newtonian effects would
have been of the same order while the tidal ones would have turned out
much smaller.
This can be considered as a positive result since, in this way,
one gets extra information about the compactness of the stars.

The system of equations that have been used in this work would be more
precise if the radiation loss due to non-radial oscillations of the
stars would have been included. Nevertheless, to leading order,
the above analysis is still valid since the energy stored
in the stellar pulsations is a fraction of the total energy of the system
(notice that the reaction Hamiltonian, equation (19),
is proportional to the non-spherical-symmetric part of the sum of
two times the kinetic energy tensor plus the potential energy
tensor and, implicitly, to the first time derivative of that part).
Furthermore, the damping time of the stellar oscillations
which for identical stars is of about the same order as the
coalescence time (coalescence understood as the stage where the
interbody-body separation amounts to some few single-body diameters only)
becomes longer as the stars become less relativistic;
this can be proven easily from the relations given in
Balbinski and Schutz \shortcite{BS82} eq. (1) and/or Reisenegger and
Goldreich \shortcite{ReGol94} eq. (11).
In our work we have taken the most compact star as a point-like mass
while the other star which is less relativistic was the one that has undergone
oscillations, oscillations with indeed much longer damping time
than the coalescence time of the binary system.
But if the evolution of a binary system is studied from the regime of
10Hz, since the coalescence time is long, the damping of the stellar
oscillations should be taken into account.

\subsection*{Acknowledgements}
KDK acknowledges the kind hospitality and the financial support
provided by the Max-Planck-Research-Group Gravitationstheorie at the
University of Jena. Helpful suggestions and discussions with Bernard
Schutz are also gratefully acknowledged by KDK.
We are grateful to Andrzej Krolak
for his verification for some of our numerical results.
Finally, we express our gratitude to an anonymous referee for
his suggestions which improved the final form of the paper.

\newpage

\section*{Tables}

\qquad {\bf Table 1:}
For $\ell = 2$, nine polytropic neutron stars models are listed for three
polytropic indices $N=0.5$, $N=1$ and $N=2$.
The index on the letters A, B, C shows the polytropic index.
For these models, which span a wide range of possible
neutron stars radii, masses, and densities, the $f$-mode frequency
is given in units of Hz.

{\bf Table 2:}
For $\ell = 2$,
normalized frequencies for three polytropic-adiabatic index pairs are listed
together with the corresponding values for the overlap integral $Q_{n2}$.

\newpage

\tt
\begin{tabular}{|c|c|c|c|c|c|c|}
\multicolumn{6}{c} {\bf Table 1} \\ \hline
\multicolumn{6}{|c|}{NEUTRON STAR MODELS} \\ \hline \hline
 Model    &  r (Km) & m (Km)     &  m/M$_{\odot}$ & $\rho_c$ (gr/cm$^3$)
 &$f$-mode (Hz)
 \\ \hline \hline
$A_{0.5} $    & 10.105  &  2.036
   &  1.38    & $ 1.17\times 10^{15} $  & 2212.8 \\
$B_{0.5} $    & 12.533  &  1.561
   &  1.06    & $ 4.69\times 10^{14} $  & 1402.4 \\
$C_{0.5} $    & 18.162  &  1.144
   &  0.77    & $ 1.13\times 10^{14} $  &  688.4
 \\ \hline \hline
$A_1 $    & 10.105  &  2.036
   &  1.38    & $ 2.09\times 10^{15} $  & 2607.6 \\
$B_1 $    & 12.533  &  1.561
   &  1.06    & $ 8.40\times 10^{14} $  & 1324.8 \\
$C_1 $    & 18.162  &  1.144
   &  0.77    & $ 2.02\times 10^{14} $  &  809.9
\\ \hline \hline
$A_2 $    & 10.105  &  2.036     &  1.38
   & $ 7.25\times 10^{15} $  & 3742.6 \\
$B_2 $    & 12.533  &  1.561     &  1.06
   & $ 2.91\times 10^{15} $  & 2372.0 \\
$C_2 $    & 18.162  &  1.144     &  0.77
   & $ 7.02\times 10^{14} $  & 1164.3 \\ \hline
\end{tabular}

\newpage

\begin{tabular}{|c|c|c|c|c|}
\multicolumn{5}{c} {\bf Table 2a} \\
\hline
\multicolumn{5}{|c|}{Eigenfrequencies and Overlap Integrals} \\ \hline \hline
\multicolumn{5}{|c|}{N = 0.5, $\Gamma_1=3.05$\qquad \qquad \qquad \qquad
N = 1, $\Gamma_1=2.05$} \\ \hline \hline

Mode  &  $\omega_{n2}^2  [G m/r^3]$   &  $|Q_{n2}|$
& $\omega_{n2}^2 [G m/r^3]$ &
$|Q_{n2}|$    \\ \hline
 $ p_2 $ & 48.2425  & .1155$ \times 10^{-3} $
& 30.1934  & .2541$ \times 10^{-2}$ \\
 $ p_1 $ & 18.8887  & .7475$ \times 10^{-2} $
& 12.3929  & .2588$ \times 10^{-1}$ \\
 $ f   $ &  1.0883  & .6238$ \times 10^{+0} $
&  1.5064  & .5580$ \times 10^{+0}$ \\
 $ g_1 $ &  0.0088  & .3193$ \times 10^{-3} $
&  0.0340  & .1764$ \times 10^{-2}$ \\
 $ g_2 $ &  0.0038  & .5054$ \times 10^{-4} $
&  0.0161  & .4250$ \times 10^{-3}$ \\
 $ g_3 $ &  0.0022  & .3106$ \times 10^{-4} $
&  0.0095  & .1256$ \times 10^{-3}$ \\
\hline
\end{tabular}

\bigskip
\bigskip

\begin{tabular}{|c|c|c|c|c|}
\multicolumn{5}{c} {\bf Table 2b}\\
\hline
\multicolumn{5}{|c|}{Eigenfrequencies and Overlap Integrals} \\ \hline \hline
\multicolumn{5}{|c|}{N = 1, $\Gamma_1=7/3$\qquad \qquad \qquad \qquad  N = 2,
$\Gamma_1=5/3$} \\ \hline \hline

Mode  &  $\omega_{n2}^2  [G m/r^3]$   &  $|Q_{n2}|$
& $\omega_{n2}^2 [G m/r^3]$ &
$|Q_{n2}|$    \\ \hline
&   61.1329  &   .3095$ \times 10^{-2}$ \\
&   40.6315  &   .8070$ \times 10^{-2}$ \\
 $   p_2 $ &  35.1837  &    .2204$ \times 10^{-2} $
&   24.0741  &   .2287$ \times 10^{-1}$ \\
 $   p_1 $ &  14.7537  &    .2121$ \times 10^{-1} $
&   11.5549  &   .7574$ \times 10^{-1}$ \\
 $   f   $ &   1.5112  &    .5580$ \times 10^{+0} $
&    3.1133  &   .4219$ \times 10^{+0}$ \\
 $   g_1 $ &   0.1902  &    .1110$ \times 10^{-1} $
&    0.5633  &   .2105$ \times 10^{-1}$ \\
 $   g_2 $ &   0.0921  &    .2607$ \times 10^{-2} $
&    0.2967  &   .8596$ \times 10^{-2}$ \\
 $   g_3 $ &   0.0549  &    .7634$ \times 10^{-3} $
&    0.1839  &   .3844$ \times 10^{-2}$ \\
&    0.1254  &   .1821$ \times 10^{-2}$ \\
&    0.0911  &   .8946$ \times 10^{-3}$ \\
&    0.0693  &   .4494$ \times 10^{-3}$ \\ \hline
\hline
\end{tabular}
\rm

\newpage
\section*{Figures}

\qquad {\bf Figure 1:} The evolution of the separation of a binary system
is shown. The system is left to evolve from a quite small
separation of $24 M$ or $125 Hz$ orbital frequency.
The dash-dotted and dashed
lines represent the evolution as determined from the
Newtonian and 1PN approximation, respectively, each time
including the appropriate radiation reaction terms, i.e. the
2.5PN approximation for the Newtonian motion and 3.5PN approximation for the
1PN motion.
The continuous line corresponds to
the evolution of the system as dictated only from the Newtonian dynamics
together with radiation reaction and tidal interaction. Slightly
inconsistently, 2.5PN (or 3.5PN)
means only the radiation reaction part without inclusion of the lower
`integer' approximations.

{\bf Figure 2:} The difference of the separation of the stars is
shown as function of time considering, additionally to the Keplerian motion,
radiation reaction on the one side and
radiation reaction + tidal interaction on the other side.
The figure presents the effects of the $f$-, $g_1$- and $g_2$-mode of star 1
(model C$_{0.5}$) with its corresponding frequencies of
about 688.4, 61.8 and 40.9 Hz.
The contributions of the analogous modes of star 2 (model A$_{0.5}$)
are negligible. Time is plotted in terms of $R = R_{rad}$.

{\bf Figure 3:} The difference of the separation of the stars is
shown as function of time considering, additionally to the Keplerian motion,
radiation reaction on the one side and
radiation reaction + tidal interaction on the other side.
The figure presents the effects of the $f$-, $g_1$-, and $g_2$-mode of star 1
(model C$_1$) with its corresponding frequencies of
809.9, 121.7 and  83.8 Hz, respectively.
The contributions from the analogous modes of star 2 (model A$_1$)
are negligible.
The $(m=2)$-resonances for the $g_1$ and $g_2$
occur near $R/M = 39.3$ and $50.5$, respectively.
The resonance of the $g_1$-mode can be clearly seen in this Figure.
Time is plotted in terms of $R = R_{rad}$.

{\bf Figure 4:} For $N=0.5$ and $N=1$, the phase differences between
the evolution
predicted by Newtonian and 1PN approximation are shown.
The tidal effects slowly accumulate a phase difference which becomes large
only at the final stage of the coalescence leading there to a phase difference
of a few complete cycles.

{\bf Figure 5:} The evolution of the separation of the two stars due
to tidal effects {\it only} (radiation reaction is not included).
For each system there is a characteristic distance beyond which the
system is driven into coalescence without the need of radiation reaction.

\end{document}